\documentclass[aps,prl,twocolumn,showkeys,superscriptaddress]{revtex4-1}

\usepackage{graphicx}

\begin{document}

\title{
  First elastic electron scattering
  from $^{132}$Xe
  at the SCRIT facility
}


\author{K.~Tsukada}
\affiliation{Research Center for Electron Photon Science, Tohoku University, Sendai, Miyagi 982-0826, Japan}
\affiliation{RIKEN Nishina Center, Wako, Saitama 351-0198, Japan}
\author{A.~Enokizono}
\affiliation{Department of Physics, Rikkyo University, Toshima, Tokyo, Japan}
\affiliation{RIKEN Nishina Center, Wako, Saitama 351-0198, Japan}
\author{T.~Ohnishi}
\affiliation{RIKEN Nishina Center, Wako, Saitama 351-0198, Japan}
\author{K.~Adachi}
\affiliation{Department of Physics, Rikkyo University, Toshima, Tokyo, Japan}
\author{T.~Fujita}
\affiliation{Department of Physics, Rikkyo University, Toshima, Tokyo, Japan}
\author{M.~Hara}
\affiliation{RIKEN Nishina Center, Wako, Saitama 351-0198, Japan}
\author{M.~Hori}
\affiliation{Department of Physics, Rikkyo University, Toshima, Tokyo, Japan}
\author{T.~Hori}
\affiliation{RIKEN Nishina Center, Wako, Saitama 351-0198, Japan}
\author{S.~Ichikawa}
\affiliation{RIKEN Nishina Center, Wako, Saitama 351-0198, Japan}
\author{K.~Kurita}
\affiliation{Department of Physics, Rikkyo University, Toshima, Tokyo, Japan}
\author{M.~Matsuda}
\affiliation{Research Center for Electron Photon Science, Tohoku University, Sendai, Miyagi 982-0826, Japan}
\author{T.~Suda}
\affiliation{Research Center for Electron Photon Science, Tohoku University, Sendai, Miyagi 982-0826, Japan}
\affiliation{RIKEN Nishina Center, Wako, Saitama 351-0198, Japan}
\author{T.~Tamae}
\affiliation{Research Center for Electron Photon Science, Tohoku University, Sendai, Miyagi 982-0826, Japan}
\affiliation{RIKEN Nishina Center, Wako, Saitama 351-0198, Japan}
\author{M.~Togasaki}
\affiliation{Department of Physics, Rikkyo University, Toshima, Tokyo, Japan}
\author{M.~Wakasugi}
\affiliation{RIKEN Nishina Center, Wako, Saitama 351-0198, Japan}
\author{M.~Watanabe}
\affiliation{RIKEN Nishina Center, Wako, Saitama 351-0198, Japan}
\author{K.~Yamada}
\affiliation{Department of Physics, Rikkyo University, Toshima, Tokyo, Japan}

\date{\today}

\begin{abstract}
  The first elastic electron scattering has been successfully performed
  at the self-confining RI ion target (SCRIT) facility,
  the world's first electron scattering facility for exotic nuclei.
  The SCRIT technique achieved high luminosity
  (over 10$^{27}$~cm$^{-2}$s$^{-1}$, sufficient for determining the nuclear shape)
  with only 10$^8$ target ions.
  While $^{132}$Xe used in this time as a target is stable isotope,
  the charge density distribution was firstly extracted 
  from the momentum transfer distributions of the scattered electrons by comparing the results with 
  those calculated by a phase shift calculation. 
\end{abstract}


\maketitle



The charge density distribution of the nucleus is one of the most important
factors in the nuclear structure investigations,
as it directly relates to the superimposition of the squared wave functions of all protons in the nucleus.
Following the monumental measurements by R.~Hofstadter and his colleagues~\cite{Hofstadter:1956}
in the latter half of the 20th century,
many stable nuclei have been studied 
by elastic electron scattering experiments.
However, with few exceptions, electron scattering from short-lived unstable nuclei
has been precluded 
by the difficulty in preparing the target material for these nuclei.
Realizing electron scattering for unstable nuclei has been long waited,
as it has been revealed that some of nuclei far from the stability valley exhibit exotic features
such as neutron halo, neutron skin, etc.~\cite{Tanihata:2013}
which are totally unknown in stable nuclei.

We have invented an internal target-forming technique
called self-confining RI ion target~(SCRIT)~\cite{Wakasugi:2004aa} in an electron storage ring,
which three-dimensionally traps the target ions along the electron beam axis.
The ions are confined by transverse focusing force given by the electron beam itself
and an electrostatic potential well provided by electrodes put along the beam axis.
After a successful feasibility study~\cite{Wakasugi:2008aa,Suda:2009zz},
we have recently completed the construction of the SCRIT electron scattering facility~\cite{Wakasugi:2013nja} 
at RIKEN's RI Beam Factory,
which is dedicated to the study of exotic nuclei.
The luminosity required for elastic electron scattering (10$^{27}$~cm$^{-2}$s$^{-1}$)
was achieved
with only 10$^8$ target ions
as available at an conventional isotope separation on line (ISOL) facility.
In traditional electron scattering experiments,
the number of target nuclei is typically of the order of 10$^{20}$.
This advancement enables electron scattering
not only from unstable nuclei, but also from stable nuclei that have not been studied to date.

In this Letter,
we report the first elastic electron scattering results of $^{132}$Xe nuclei obtained at the SCRIT facility.
Although $^{132}$Xe is a stable nucleus, it has never been investigated by electron scattering~\cite{deVries:1987}.
Interestingly,
stable xenon isotopes have been recently utilized as targets for dark matter searches~\cite{Teresa:2016,PandaX-II:2016,LUX:2017}, 
and in neutrinoless double beta decay experiments~\cite{Frank:2008}.
To calculate the cross sections in these experiments,
the form factors of the Xe isotopes are required.
However,
transition X-ray measurements of muonic atoms have yielded 
only the root-mean-squares of the Xe isotopic charge radii~\cite{Fricke:1995}.
In the present study,
we newly determine the surface shape of the $^{132}$Xe nucleus in the Xe isotopes.

\begin{figure*}[hbt]
\includegraphics[width=13cm]{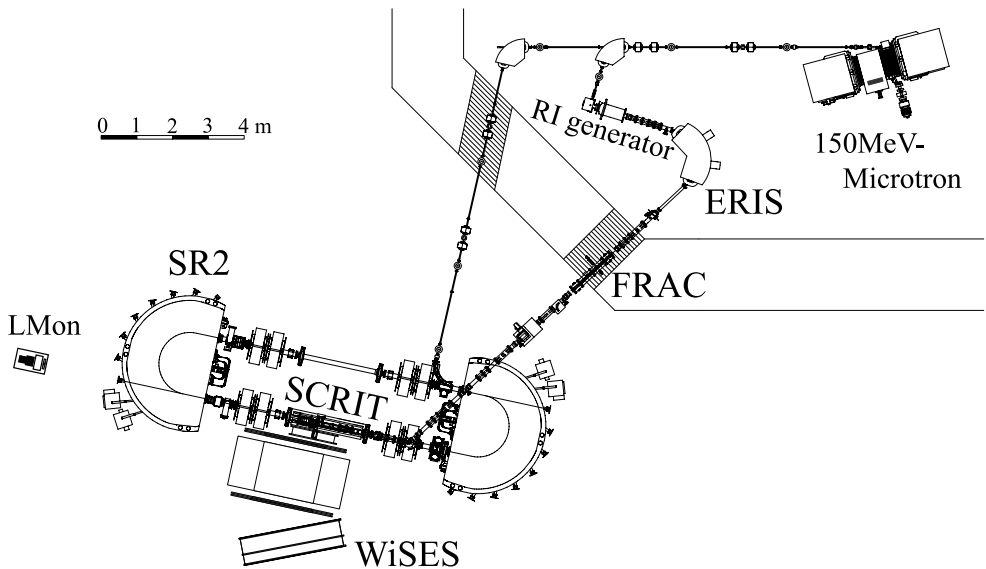}%
\caption{\label{Fig:SCRITfacility}
  Overview of the SCRIT electron scattering facility.
}
\end{figure*}

Figure~\ref{Fig:SCRITfacility} illustrates the SCRIT facility.
Details of the facility are described elsewhere~\cite{Suda:2012mi,Wakasugi:2013nja}.
The 150~MeV electron beam is injected and accumulated in the storage ring (SR2).
The electron beam is then accelerated to the desired energy below 700~MeV if necessary.
The Xe ions are produced from natural Xe gas, and accelerated and mass-separated by functions of an electron-beam driven RI separator for SCRIT (ERIS)~\cite{Ohnishi:2013bia}.
Continuous ion beam from the ERIS is converted to pulsed beam with the duration of 370~$\mu$s at a fringing-RF-field-activated ion beam compressor (FRAC)~\cite{Togasaki:2015}
and it is delivered to the SCRIT.
The scattered electrons from confined ions in the SCRIT are then detected by a window-frame spectrometer for electron scattering (WiSES).
The luminosity is experimentally determined by counting the bremsstrahlung photons (produced by collisions with target ions) with a luminosity monitor (LMon) put at 8-m downstream from the SCRIT.

The SCRIT device is composed of four cylindrical electrodes
surrounding the electron beam.
The middle two electrodes define the trapping length of 50~cm in total,
and they are separated with 1-cm gap to provide a space for insertion of a tungsten wire,
which was used for the study of the spectrometer acceptance.
These electrodes form an electrostatic potential lower than the acceleration voltage of delivered target ions by a few V
to make the ion motion slow and to make confinement time longer.
Electrodes at both ends make barrier potentials for the longitudinal confinement.

The $^{132}$Xe ions were trapped in the SCRIT for 240~ms,
then ejected to refresh the target quality,
although the confinement lifetime for target ions is typically 2--3~s.
The ejection was implemented by controlling one of the gate potentials on the SCRIT device.
Ejection reduces the contribution of residual gas which becomes ionized and trapped by the electron beam,
and hence accumulates over time.
To estimate the background (contributed mainly by the residual gases),
the injection-trapping-ejection sequences was alternately repeated with and without the target ions
at 10-ms intervals.
The number of $^{132}$Xe ions introduced to the SCRIT with each cycle was less than a few times 10$^{8}$.

The WiSES spectrometer consists of a dipole magnet, 
drift chambers at the entrance and exit of the magnet,
two scintillation counters for trigger generation,
and
a helium-gas filled bag constructed from a 30~$\mu$m-thick vinyl.
The bag is installed between the two drift chambers 
to reduce the multiple-scattering effect.
The spectrometer magnet is a window-frame dipole magnet with a large aperture.
Its dimensions are 29~cm~(H), 171~cm~(W), 140~cm~(D),
and its magnetic field is uniform except at the inlet and outlet.
The trajectories of the scattered electrons are reconstructed using
a three-dimensional field map calculated by a finite element method (TOSCA~\cite{TOSCA}).
The calculated map was confirmed to well-reproduce the vertical component of the magnetic field measured with a hall probe.
During measurements,
the magnitude of the magnetic field was monitored
by an NMR probe positioned in the homogeneous field region.
The solid angle of the spectrometer is approximately 80~msr 
covering scattering angles from 30 to 60$^{\circ}$.
For the fixed position and opening angle of the WiSES,
the electron beam energy ($E_e$) was varied as 151, 201, and 301~MeV,
covering the momentum transfer region 0.4--1.5~fm$^{-1}$.
The magnetic field of WiSES was adjusted to 0.41, 0.54, and 0.80~T correspondingly,
and the momentum resolutions ($\delta p/p$) evaluated in the simulation was
3.7$\times$10$^{-3}$, 2.8$\times$10$^{-3}$, and 2.0$\times$10$^{-3}$ respectively.
At the beginning of the measurement,
the accumulated electron beam current was typically 250~mA.
The beam had interacted with target ions and residual gases in the storage ring,
which reduced its current to 150~mA
at the end of the data-taking.
A typical beam size was 2~mm$^H\times$1~mm$^V$ ($\sigma$) at the center of the SCRIT.

\begin{figure}[tbh]
  \includegraphics[width=8cm]{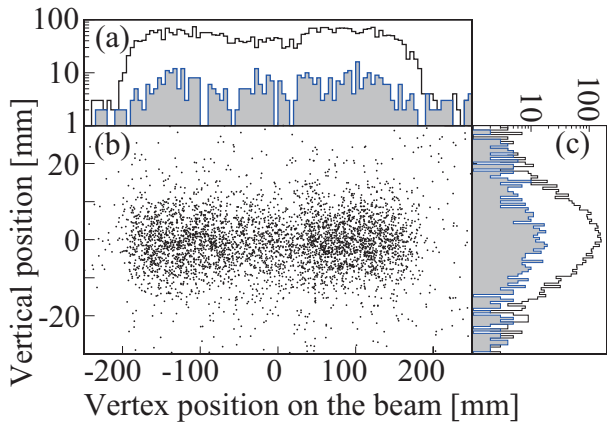}%
  \caption{\label{Fig:VyVz}
    Reconstructed vertex distributions of the $^{132}$Xe target at $E_e$~=~151~MeV.
    The low-energy background is rejected by the momentum selection.
    The top and bottom of the SCRIT electrodes are placed at $\pm$20~mm in vertical position.
    Panels (a) and (c) show the vertex point distributions with and without the target ions (plain and hatched histograms, respectively).
  }
\end{figure}

Panels (a)--(c) of Fig.~\ref{Fig:VyVz} show reconstructed vertex distributions
along the beam and at vertical positions
after removing the low-energy background at $E_e$~=~151~MeV.
Target ions were clearly trapped along the beam-line between the top and bottom electrodes of the SCRIT
put at the $\pm$20~mm in the vertical positions.
Since the barrier potentials are leaky,
the effective longitudinal trapping region was shorter than the electrodes (40~cm vs. 50~cm).
The depletion at the center of Fig.~\ref{Fig:VyVz}~(a) was formed by that
highly ionized ions were localized at two shallow potential minimums
due to the gap at the center of the SCRIT.
The width of the vertical distribution was 6.3~mm ($\sigma$),
consistent with the vertical position resolution
evaluated by using the wire target.
The shaded histograms in Fig.~\ref{Fig:VyVz}~(a) and (c) are the background contributions 
from residual gases measured
in the absence of target ions.
These contributions were approximately estimated as 10\%, 30\%, and 40\% 
for $E_e$~=~151, 201, and 301~MeV, respectively.
Assuming $^{16}$O as the residual gas,
the approximate background luminosity was estimated as 1$\times$10$^{27}$~cm$^{-2}$s$^{-1}$,
roughly consistent with the estimation from vacuum pressure ($\sim$5$\times$10$^{-8}$~Pa around the SCRIT region).

\begin{figure}[tbh]
  \includegraphics[width=8cm]{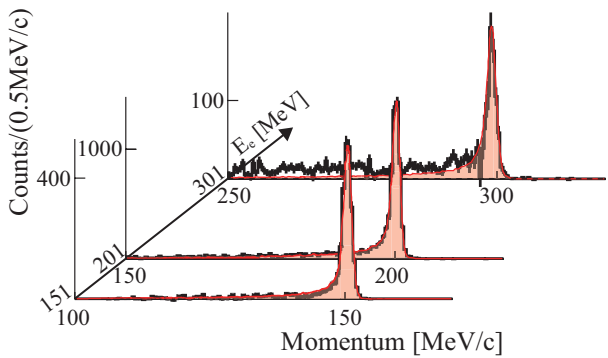}%
  \caption{\label{Fig:MomDistXe}
    Reconstructed momentum spectra of $^{132}$Xe target after background subtraction.
    Red shaded lines are the simulated radiation tails following the elastic peaks.
  }
\end{figure}

Figure~\ref{Fig:MomDistXe} shows the reconstructed momentum spectra
at $E_e$~=~151, 201, and 301~MeV,
obtained after subtracting the background.
Elastic events clearly manifest as peaks in the spectra.
The measured
$\delta p/p$ were $\sim$5.4$\times$10$^{-3}$, 3.7$\times$10$^{-3}$, and 3.0$\times$10$^{-3}$,
respectively.
The present momentum resolutions are slightly below the design values.
Possible reasons for this discrepancy are the imperfection of knowledge in the magnetic field of the spectrometer,
a small amount of air contamination in the helium bag,
and the energy spread of the electrons circulating in the SR2.
As shown in the figure, the low-energy tails below the elastic peak at $E_e$~=~151~MeV and 201~MeV 
were well-reproduced by simulations
of a well-known radiative process~\cite{Friedrich:1975}.
At $E_e$~=~301~MeV, 
the enhanced tail suggests some inelastic processes.
In the high-momentum transfer region (0.9--1.4~fm$^{-1}$),
the magnitude of elastic scattering diminishes
and inelastic scattering processes (such as the giant dipole resonance) gain prominence.

\begin{figure}[t]
  \includegraphics[width=8cm]{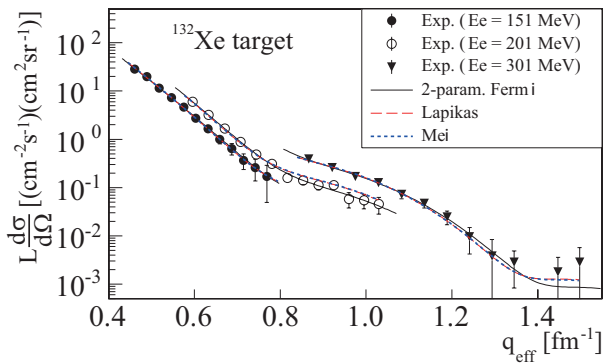}%
  \caption{\label{Fig:LCS_Xeon}
    Differential cross sections multiplied by luminosity, versus effective momentum transfer
    for $E_e$~=~151~MeV (black filled circles), 201~MeV (open circles) and 301~MeV (filled triangles).
    The lines are the results of DWBA calculations
    assuming
    nuclear charge density distributions obtained by
    the two-parameter Fermi distribution (black solid line),
    the Hartree-Fock~+~phenomenological calculation (red dashed line)~\cite{Lapikas:2004},
    and the beyond-relativistic-mean-field theory (blue dotted line)~\cite{Hagino:2016}.
    The parameters of the two-parameter Fermi distribution are best values evaluated from Fig.\ref{Fig:CtFit}.
  }
\end{figure}

Figure~\ref{Fig:LCS_Xeon} shows differential cross sections
of the $^{132}$Xe
multiplied by luminosity,
as functions of effective momentum transfer ($q_{eff}$)
for $E_e$~=~151, 201 and 301~MeV.
The $q_{eff}$, which accounts for the Coulombic attraction between electrons and nuclei,
is defined as,
$q_{eff}=q\left( 1 + \frac{3}{2}\left( Z\alpha/E_{i}R \right) \right),
R=1.2\times A^{1/3}$~fm~\cite{Uberall}.
In this expression,
$q$ is the momentum transfer calculated from the measured angle as $q=2E_i\sin(\theta/2)$,
$E_{i}$ is the initial electron energy,
and $\theta$ is the polar angle of the scattered electrons.
$Z$ and $A$ are the atomic and mass numbers of the nucleus respectively,
and $\alpha$ is the fine structure constant.
The systematic error in the cross section,
introduced by ambiguity in the spectrometer acceptance,
was estimated as approximately 5\%.

\begin{figure}[t]
  \includegraphics[width=8cm]{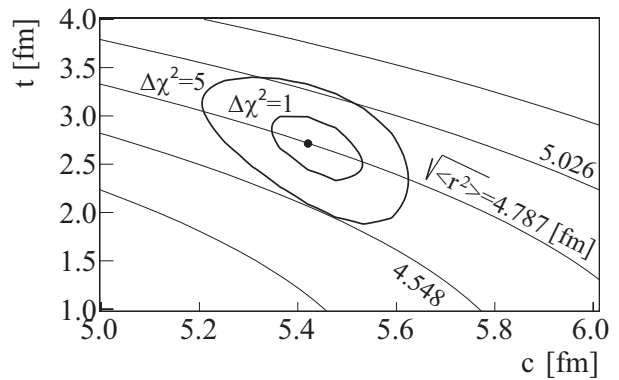}%
  \caption{\label{Fig:CtFit}
    Contour $\chi^{2}$ plot of parameter determination,
    obtained by fitting to
    DWBA calculations with different nuclear shape parameters $c$
    and $t$.
    The three data sets ($E_e$~=~151, 201, and 301~MeV) are fitted simultaneously.
    Slanted lines are curves of constant root-mean-square charge radii.
    The 4.787~fm line represents the result of X-ray measurements of muonic atoms~\cite{Fricke:1995}.
  }
\end{figure}

The lines in Fig.~\ref{Fig:LCS_Xeon} are the elastic scattering cross sections calculated 
by a phase shift calculation code DREPHA~\cite{drepha}
with a nuclear charge density distribution.
The solid line assumes a two-parameter Fermi distribution:
$\rho(r)=\rho_{0}/(1+\exp{(4.4(r-c)/t)})$,
where
$\rho_{0}$ is the density at the center of the nucleus,
and $c$ and $t$ are surface distribution parameters.
In the present analysis, the luminosity at each energy was considered as a fitting parameter.
Determination of the absolute luminosity by LMon is currently underway.
The luminosities evaluated by fitting the two-parameter Fermi distribution were 0.87$\times$10$^{27}$, 1.06$\times$10$^{27}$, and 1.55$\times$10$^{27}$~cm$^{-2}$s$^{-1}$ for $E_e$~=~151, 201, and 301~MeV, respectively,
in which best values of the parameters $c$ and $t$ are used as evaluated later (see Fig.~\ref{Fig:CtFit}).
The dashed line is the calculation with the Hartree-Fock calculation proposed by Lapik\'{a}s~\cite{Lapikas:2004},
which phenomenologically incorporates the shell effect
by interpolation between the Hartree-Fock calculations of neighboring nuclei, $^{124}$Sn and $^{138}$Ba.
The dotted line is calculated by using the beyond-relativistic-mean-field theory in Mei~\cite{Mei:2015,Hagino:2016}.
In the both calculations,
only the luminosities were considered as fitting parameters
because the nuclear shapes are fixed.
In general,
the both theoretical calculations agree with our data within the experimental errors.
The experimental data are expected to be contaminated with inelastic scattering processes corresponding to some excited states (e.g., $2^+$, excitation energies are 0.67~MeV, 1.3~MeV and higher),
because the momentum resolution of our spectrometer 
is insufficient for separating these states from the ground state.
Theoretical calculations
with a distorted wave Born approximation code FOUBES~\cite{Heisenberg:1981,Foubis:1983}
using a transition density distribution~\cite{Hagino:2016} 
indicate
that below 1.2~fm$^{-1}$ and at 1.4~fm$^{-1}$,
the first excited state contributes
below 1\% and 10\% of the ground-state contribution, respectively.
These contributions are much smaller than the statistical errors.

Figure~\ref{Fig:CtFit} shows the contours of fitting $\chi^{2}$
for different $c$ and $t$ in the two-parameter Fermi distribution.
The contour levels of the root-mean-squares charge radii at each $c$ and $t$ are also shown (slanted lines).
The innermost contour level represents the $\Delta\chi^{2}<$1 region,
which yields 
the most probable values ($c$~=~5.4$^{+0.1}_{-0.1}$~fm, $t$~=~2.7$^{+0.3}_{-0.4}$~fm, and $\left<r^2\right>^{1/2}$~=~4.8$^{+0.1}_{-0.1}$~fm).
The large error in the diffuseness in the present analysis is due to the absence of the experimentally determined absolute value of the luminosity.
This deficiency will be improved by the LMon development.

Our calculated root-mean-square charge radius is consistent with
that obtained by X-ray measurements of muonic atoms, namely, $\left<r^2\right>^{1/2}$~=~4.787~fm~\cite{Fricke:1995}.
It is worth noting that the theoretical root-mean-square charge radii reported by Lapik\'{a}s~\cite{Lapikas:2004}
and Mei~\cite{Hagino:2016} (4.804~fm and 4.806~fm respectively) also lie within the error bounds of our result.

In conclusion,
we extracted
information on the nuclear shape of $^{132}$Xe
by measuring the elastic electron scattering from $^{132}$Xe at the SCRIT electron scattering facility.
The momentum transfer distributions of the differential cross sections are mostly consistent with theoretical calculations;
especially, the root-mean-square charge radii agree within the experimental errors.
Assuming the two-parameter Fermi distribution model as the nuclear charge distribution,
the shape parameters were determined as $c$~=~5.4$^{+0.1}_{-0.1}$~fm, and $t$~=~2.7$^{+0.3}_{-0.4}$~fm.
This work demonstrates that the SCRIT technique enables up to perform electron scattering experiment for determination of nuclear charge density distribution even for small amount of isotopes provided by conventional ISOL system.

RI production for experiments on unstable nuclei has already started~\cite{Ohnishi:2015qhb}.
Electron scattering off short-lived unstable nuclei will be realized in the near future.

The authors would like to thank 
H.P.~Blok, L.~Lapik\'{a}s, K.~Hagino and H.~Mei for critical supports of calculations for nuclear charge density distributions and electron scattering cross sections.
This work has been supported by Grants-in-Aid for Scientific Research (S) (Grands No. 22224004) from JPS.

\bibliography{SCRIT}

\end{document}